\documentclass[sigconf]{acmart}

\copyrightyear{2023}
\acmYear{2023}
\setcopyright{acmlicensed}\acmConference[HILDA '23]{Workshop on Human-In-the-Loop Data Analytics}{June 18, 2023}{Seattle, WA, USA}
\acmBooktitle{Workshop on Human-In-the-Loop Data Analytics (HILDA '23), June 18, 2023, Seattle, WA, USA}
\acmPrice{15.00}
\acmDOI{10.1145/3597465.3605224}
\acmISBN{979-8-4007-0216-7/23/06}

\usepackage[english]{babel}
\usepackage{caption}
\usepackage{subcaption}
\usepackage{algpseudocode}
\usepackage{lscape}
\usepackage[linesnumbered,ruled,vlined]{algorithm2e}
\usepackage{amsmath}
\usepackage{graphicx}
\usepackage{cleveref}
\usepackage{enumitem}
\usepackage{adjustbox}

\usepackage{multirow}

\usepackage{listings}

\usepackage{colortbl}

\usepackage{tikz}
\usetikzlibrary{shapes.arrows}

\newtheorem{theo}{Theorem}

\newtheorem{example}[theo]{Example}

\theoremstyle{remark}

\usepackage{dsfont}

\definecolor{darkgreen}{rgb}{0.15,0.55,0.15}
\definecolor{darkblue}{rgb}{0.1,0.1,0.5}
\definecolor{blue}{rgb}{0.01,0.40,.8}
\definecolor{darkgreen}{rgb}{0.15,0.55,0.15}
\definecolor{mred}{rgb}{.80,.12,.30}
\definecolor{grey}{rgb}{0.5,0.5,0.5}
\definecolor{Purple}{rgb}{.75,0,.85}
\definecolor{light-gray}{gray}{0.95}
\definecolor{mid-gray}{gray}{0.85}
\definecolor{darkred}{rgb}{0.7,0.25,0.25}
\definecolor{rose}{rgb}{1.0, 0.01, 0.24}

\newcommand{\gray}[1]{\textcolor{grey}{#1}}

\usepackage{latexsym}

\newcommand{\eat}[1]{}

\newcommand{\stitle}[1]{\vspace{2pt}\noindent\textbf{#1}}

\author{Zezhou Huang}
\email{zh2408@columbia.edu}
\affiliation{
  \institution{Columbia University}
}

\author{Eugene Wu}
\email{ewu@cs.columbia.edu}
\affiliation{
  \institution{DSI, Columbia University}
}

\begin{document}

\title{Cocoon: Semantic Table Profiling Using Large Language Models}

\begin{abstract}
Data profilers play a crucial role in the preprocessing phase of data analysis by identifying quality issues such as missing, extreme, or erroneous values. Traditionally, profilers have relied solely on statistical methods, which lead to high false positives and false negatives. For example, they may incorrectly flag missing values where such absences are expected and normal based on the data's semantic context. To address these, we introduce Cocoon, a data profiling system that integrates LLMs to imbue statistical profiling with semantics. Cocoon enhances traditional profiling methods by adding a three-step process: Semantic Context, Semantic Profile, and Semantic Review. Our user studies show that Cocoon is highly effective at accurately discerning whether anomalies are genuine errors requiring correction or acceptable variations based on the semantics for real-world datasets.
\end{abstract}

\begin{teaserfigure}
  \includegraphics[width=\textwidth]{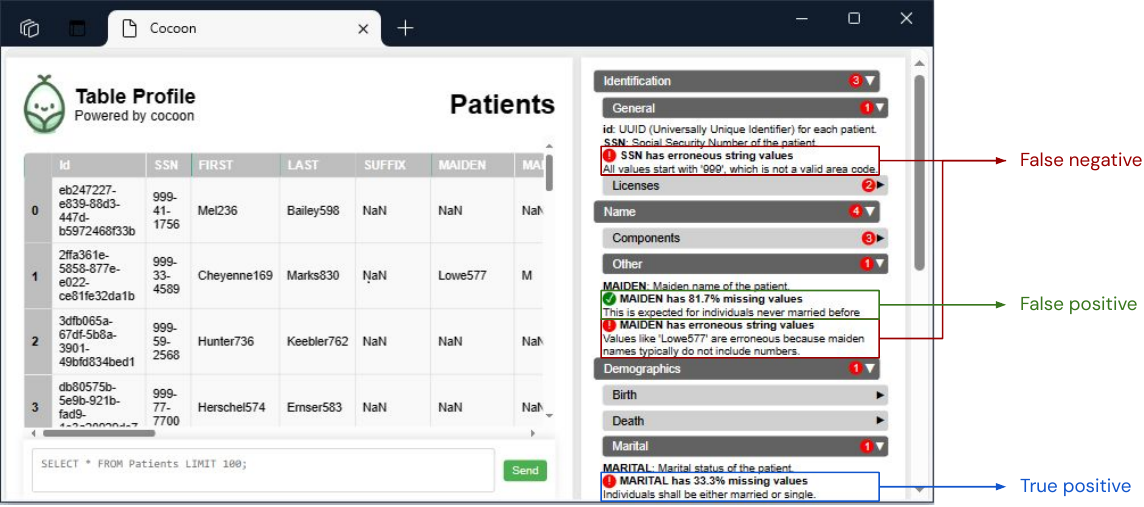}
  \caption{
Screenshot of the Cocoon table profile, showing how it reduces false positives and false negatives from earlier statistical profiles. The left panel displays a table and an SQL query box for exploration. The right panel highlights alerts for data errors missed by the statistical profile (false negative) and avoids alerting for errors that are actually normal (false positive).
  }
  \label{fig:teaser}
\end{teaserfigure}

\maketitle



\section{Introduction}

Datasets often contain a miscellaneous set of problems, such as missing, extreme, or erroneous values. These issues can significantly impact the accuracy of further analysis~\cite{sambasivan2021everyone}. Consequently, analysts devote more than 80\% of their time to manually examining and cleaning the data~\cite{eckerson2002data}. To address these issues, there is growing interest in profiling the tables to help users interactively visualize and understand data quality issues~\cite{kandel2012profiler, jannah2014metareader, liu2018steering,epperson2023dead, naumann2014data}.

While profilers are meant to be proxies for detecting real-world errors, previous profilers have relied {\it solely on Statistical Profile}. For instance, to help detect errors, Profiler~\cite{kandel2012profiler} depends on rules, patterns, numerical distributions and custom functions. This purely statistical approach can lead to incorrect error detection, both in terms of false positives and false negatives. To illustrate this, consider the Synthea~\cite{walonoski2018synthea} Patient table (\Cref{fig:teaser}) as an example:

\begin{example} (1) To identify missing value errors, Statistical Profile would apply a custom function to compute the percentage of NULL for Maiden Column as 81.7\% and alert it for the user to clean, as it's $>0$.  This is a false positive because the Maiden Column is expected to be mostly missing for unmarried individuals. (2) To identify erroneous string values for SSN, Statistical Profile computes the regex shared by most of the values and highlights those that violate the regex. This leads to false negatives because all values in the SSN start with '999', but are not valid area codes and the whole column is erroneous.
\end{example}

\begin{figure}
    \centering
    \includegraphics[width=\linewidth]{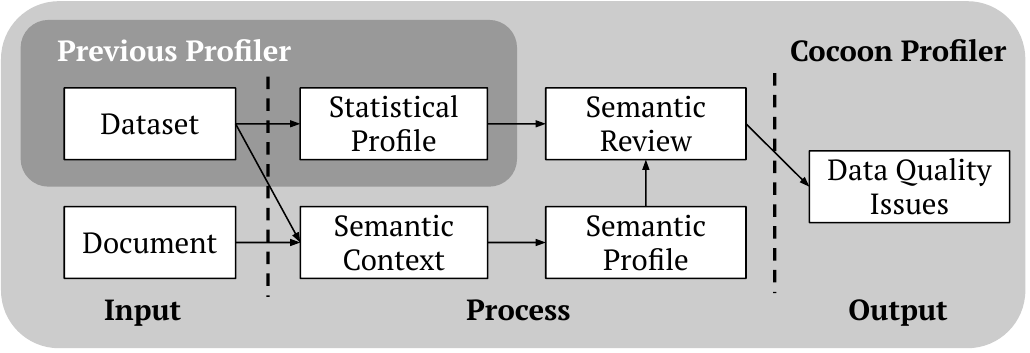}
    \caption{The process of how semantic understanding is used for profiling. Previous profiler generates only Statistical Profiles, and users have to manually  understand, profile and review whether the errors are semantically meaningful.}
    \label{fig:process}
\end{figure}

Statistical Profile alone falls short as it lacks the {\it semantics} of the tables~\cite{de2003visual,idreos2015overview, zuur2010protocol,russo2019much, huang2022reptile}—including what the tables and columns represent, and what kinds of values, patterns, and distributions are expected. These semantics are indispensable for making informed decisions about whether statistical outliers/inliers are acceptable or abnormal. In this work, we model such a process of obtaining and applying semantics to data profiles in 3 steps (\Cref{fig:process}):

\begin{enumerate}
    \item {\bf Semantic Context:}  Based on dataset samples and related documents, users gain the context of the datasets as natural language (NL) descriptions of tables and columns.

    \item {\bf Semantic Profile:} Utilizing the Semantic Context, users form expectations of what the table and columns should be, in a format similar to that of a Statistical Profile.

    \item {\bf Semantic Review:} Compare the Statistical Profile with the Semantic Profile. If there are discrepancies, assess whether these are errors or if they are semantically acceptable.

\end{enumerate} 

\begin{example} Continue the example for the Maiden Column. The (\underline{Statistical Profile}) flags it as erroneous for 81.7\% NULL values. In response, users first manually explore the table and related documents to understand that "Maiden refers to the maiden name of the patient" (\underline{Semantic Context}). Then, they realize that missing maiden names are expected (\underline{Semantic Profile}) because they are only applicable to married individuals. When they find that maiden names indeed have a large percentage of missing values, they conclude that this is normal and does not require cleaning (\underline{Semantic Review}).
\end{example}

Such a process of applying semantics to verify data quality issues extends beyond just missing values. \Cref{table:error_types} catalogs the various data quality issues studied by Profiler~\cite{kandel2012profiler}, its Statistical Profiles, and the proposed Semantic Profiles and Semantic Reviews.

However, previously, obtaining such semantics was challenging because it required a broad general knowledge of the real world. Consequently, the burden fell on users to manually sift through the datasets, understand the tables and columns, and then apply their own semantic understanding of the data and its context to interpret the alerts from the profilers.
To alleviate the burden on users, recent advancements in Large Language Models (LLMs)~\cite{bubeck2023sparks} have shown great potential as proxies for obtaining these semantics. Although these semantics are not perfect or domain-specific, and still necessitate a human-in-the-loop approach, this method eases the workload on users by requiring them only to confirm the semantics instead of generating them manually.

To this end, we present Cocoon, a data profiling system that leverages LLMs to obtain and apply the semantics to the statistically profiled errors. The core design of Cocoon is to decompose the complex tasks of semantics into three steps: Semantic Context, Semantic Profile, and Semantic Review; these steps are motivated by how humans explore data, where they first form an expectation of the data, then inspect the actual data, and finally judge the discrepancy~\cite{de2003visual, idreos2015overview, zuur2010protocol, russo2019much, huang2022reptile}.
These steps are further applied to a decomposed range of errors, including duplication, missing values, outliers, etc., as listed in \Cref{table:error_types}. Such a process of task decomposition for LLMs has been shown to be critical for the accuracy and robustness of various data tasks like visualization and transformation~\cite{khot2022decomposed, dibia2023lida, huang2024relation}. The resulting profiles generated by Cocoon are in two forms: (1) an easy-to-explore user interface and NL explanation that aids in human understanding and verification (2) a JSON format for future LLM systems to retrieve related pieces that are useful for tasks such as text-to-SQL~\cite{lewis2020retrieval, huang2023data}.

\begin{table*}[ht]
\centering
\caption{Taxonomy of error types, Statistical Profile, Semantic Profile, and the protocol of Semantic Review adopted by Cocoon.}
\label{table:error_types}
\begin{tabular}{p{2.2cm}p{4.1cm}p{3.9cm}p{6.5cm}}
\hline
\textbf{Error Type} & \textbf{Statistical Profile} & \textbf{Semantic Profile} & \textbf{Semantic Review} \\ \hline
Duplication  & \textbf{HasDuplicate} (\textit{bool}); \newline \textbf{SampleDuplicate} (\textit{table}); & \textbf{ExpDuplicate} (\textit{bool}); \newline \textbf{Thought} (\textit{str}); & If HasDuplicate, assess the SampleDuplicate acceptability. \\ \hline
Column  Type & \textbf{CurrentType} (\textit{str}); \newline \textbf{SampleValue} (\textit{list[object]}) & \textbf{ExpType} (\textit{str}); \newline \textbf{Thought} (\textit{str}); & Ensure CurrentType match ExpType. If not, assess CurrentType for acceptability. \\ \hline
Unique key & \textbf{UniqueRatio} (\textit{float}); \newline \textbf{SampleNonUnique} (\textit{list[str]}); & \textbf{ExpUnique} (\textit{bool}); \newline \textbf{Thought} (\textit{str}); & If ExpUnique and UniqueRatio $\neq1$, evaluate SampleNonUnique acceptability. \\ \hline
DMV & \textbf{CandidateDMV} (\textit{list[object]}); & \textbf{PotentialDMV} (\textit{list[object]}); \newline \textbf{Thought} (\textit{str}); & Compare CandidateDMV to PotentialDMV. Confirm CandidateDMV as DMVs or rule them out. \\ \hline
Missing Value & \textbf{MissingPercentage} (\textit{float}); \newline \textbf{SampleMissing} (\textit{table}); & \textbf{AllowMissing} (\textit{bool}); \newline \textbf{Thought} (\textit{str}); & If MissingPercentage $>0$, assess SampleMissing acceptability. \\ \hline
Numeric Outliers & \textbf{Quantile} (\textit{list[float]}); & \textbf{ExpQuantile} (\textit{list[float]}); \newline \textbf{Thought} (\textit{str}); & Ensure Quantile align with ExpQuantile. Assess deviations for acceptability. \\ \hline
String Outliers & \textbf{RegexPattern} (\textit{str}); \newline \textbf{SampleOutlier} (\textit{list[str]}); \newline \textbf{SampleInlier} (\textit{list[str]}); & \textbf{ExpStr} (\textit{list[str]}); \newline \textbf{Thought} (\textit{str}); & Validate RegexPattern, and SampleInlier against ExpStr. Assess SampleOutlier for acceptability, and SampleInlier for potential errors. \\ \hline
Missing Record & \textbf{ValueFreq} (\textit{dict[object: int]}); & \textbf{ExpFreq} (\textit{dict[object: int]}); \newline \textbf{Thought} (\textit{str}); & Check ValueFrequency alignment with ExpFreq. Assess deviations for acceptability. \\ \hline
\end{tabular}
\end{table*}

\section{Related Works}

\stitle{Error Detection.}
Previous works~\cite{mahdavi2019raha, rekatsinas2017holoclean,hellerstein2008quantitative,chu2013holistic} employ combinations of rule-based, pattern-based, or quantitative statistical approaches to (1) first learn the rules, patterns, and distribution of clean data, and then (2) detect values that violate these as errors. However, these methods depend on the input data distribution to learn the distribution of clean data~\cite{de2018formal}. Nevertheless, many real-world errors are generated systematically, for example, due to data collector malfunctions, which could render an entire column incorrect. In these cases, semantics are necessary to correctly evaluate whether inliers/outliers are errors~\cite{de2003visual,idreos2015overview, zuur2010protocol,russo2019much}.

\stitle{Data Profiler.} To obtain semantics for error detection, previous data profilers have relied on humans to manually interpret them based on their knowledge, assisted by visualizations of statistical profiles~\cite{kandel2012profiler, jannah2014metareader, liu2018steering,epperson2023dead, naumann2014data}. 
However, the challenge remains that manually sifting through these tables and columns, and interpreting these visualizations and statistics, is overwhelming for users~\cite{liu2018steering,epperson2023dead}.

\stitle{LLMs for Data Tasks.} With the advancement of LLMs, numerous works have applied LLMs to data tasks that require semantics, such as data analytics, table transformation, and visualization~\cite{ma2023insightpilot,dibia2023lida,huang2024relation}. 
Cocoon similarly utilizes LLMs for data to extract semantics for enhanced profiling. Compared to other data tasks, Cocoon can be considered an upstream system because profiling the data is a fundamental initial step before conducting further data tasks~\cite{huang2023data}.

\section{System Design}

\begin{figure}
    \centering
    \includegraphics[width=0.9\linewidth]{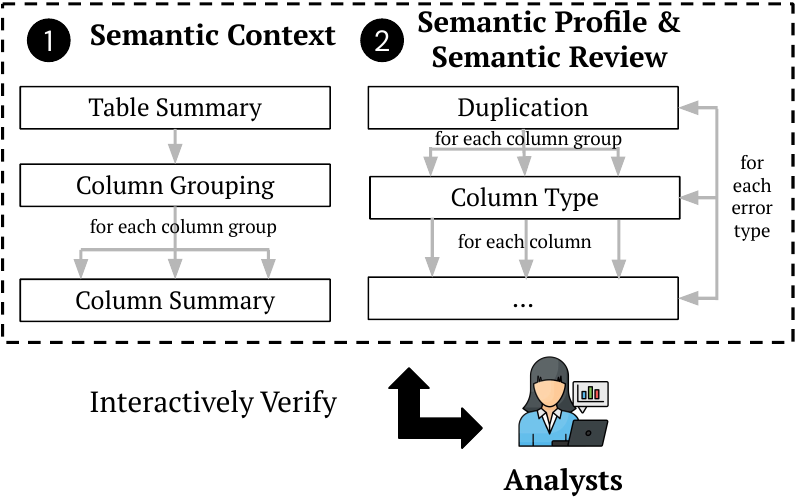}
    \caption{Cocoon System Design: First, Cocoon summarizes tables, groups attributes, and summarizes columns for Semantic Context. Then, Cocoon performs Semantic Profiling and Semantic Review for each error type, starting from the table-level error (duplication) to column-level errors. Analysts interactively verify the process.
    }
    \label{fig:arch}
\end{figure}

Data profiling involves multiple detailed steps, from a high-level understanding of the tables to the interpretation of columns, for various types of data errors. Cocoon decomposes the profiling tasks across two dimensions: the types of errors, and for each error, the steps of profiling. The scope of data errors for Cocoon follows that of previous profilers, as listed in \Cref{table:error_types}. For each type of error, Cocoon's profiling follows \Cref{fig:process}, combining statistical and semantic profiling. The final design is illustrated in \Cref{fig:arch}. We find that the Semantic Context for various errors is a reusable initial step; therefore, Cocoon performs this first. Then, Cocoon addresses each error type, starting from table-level errors (such as duplication) and then proceeding to each group of columns. The entire process is interactive, allowing users to verify and edit each step. Next, we will walk through the system design in detail.

\stitle{Setup.} 
Cocoon takes as input the database connection (currently only supports DuckDB~\cite{raasveldt2019duckdb}), the table name and LLM API key. Cocoon begins by verifying the existence of the table, and retrieving its schema. Alternatively, users can provide a Pandas DataFrame, which we register as a view using DuckDB for querying. It then uses the connection to explore data and construct LLM prompts.

\subsection{Semantic Context}

For the Semantic Context, Cocoon extracts: (1) NL summary of the table, (2) column grouping, and (3) NL summary of each column. 

\subsubsection{Table Summary}
We prompt the LLM by displaying the first five rows and, if provided, related table documents in NL for further contexts. We then instruct the LLM to summarize the high-level idea of the table in a NL description. To ensure that the summary comprehensively covers all columns, we ask the LLM to underline (\textless u\textgreater\textless /u\textgreater) all the columns to discuss how these are related. We write a Python test program to ensure all attributes are included and underlined. If it fails the tests, we retry the LLM call.

\subsubsection{Column Grouping}
We group columns for two reasons: (1) For later column-level error detection, some columns are naturally grouped together to express one semantic concept. For example, the columns "day, month, year" together express a semantically coherent date, while "longitude, latitude" express location coordinates. These columns are analyzed better together. (2) Moreover, for human understanding, some tables could have hundreds of columns, which are overwhelming to navigate. We provide a hierarchical tree view of the groups of columns to facilitate understanding.

To group the columns, we provide the table summary from the previous step and ask the LLM to generate a JSON file, which contains multiple levels of grouped concepts, with the final level being the list of attributes. We write a test program to ensure that each attribute appears in exactly one list.

\subsubsection{Column Summary}
For each column group, we provide the summary of the table and sample projections of the group of columns (10 rows) to the LLM to summarize the meaning of each column.

\subsubsection{Human Feedback}
\begin{figure}
    \centering
    \includegraphics[width=\linewidth]{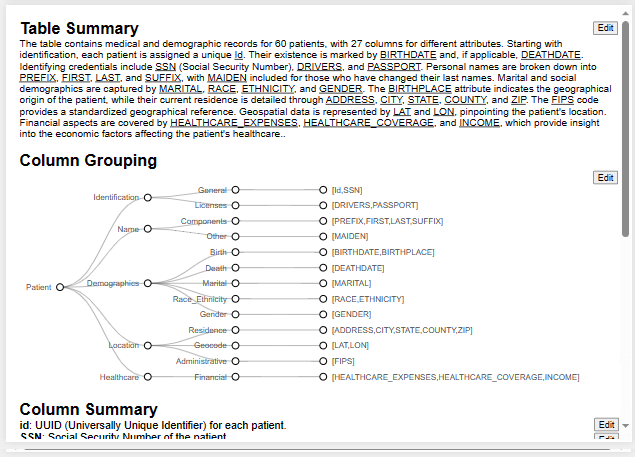}
    \caption{The panel displays the Semantic Context of the table, including the table summary, hierarchical grouping of attributes, and the column summary. Users can edit both the summary and the hierarchy.}
    \label{fig:semantic_context}
\end{figure}

For all the above Semantic Context, users can verify and edit the texts and JSON; the interface is shown in \Cref{fig:semantic_context}. We also run the same tests on table summary and column grouping  to ensure the revised outcomes are complete.

\subsection{Semantic Profile and Review}

Using the Semantic Context from the previous step, Cocoon delves into the Semantic Profile and Semantic Review for each error type, starting from the table level (Duplication) to the column group level (Column Type), and finally, to each individual column. The taxonomy of the Semantic Profile and Semantic Review is shown in \Cref{table:error_types}. Note that, for all Semantic Profiles, we ask LLMs to output their detailed reasoning thoughts as a string because (1) previous work~\cite{wei2022chain} shows that providing thoughts improves LLM performance, and (2) the thoughts are provided as context for Semantic Review. The output of the Semantic Review is a boolean indicating whether to alert an error and a string of the reasoning thoughts, for all error types.
Here, we go through each and provide examples of how they help avoid false positives and negatives.

\subsubsection{Duplications}
(1) For the Statistical Profile, Cocoon issues an SQL query to test for duplicates (\textit{HasDuplicate}). If \textit{HasDuplicate} is true, it provides a sample of up to 5 duplicated rows along with their duplication count (\textit{SampleDuplicate}), by grouping by all columns and computing the count.
(2) For the Semantic Profile, Cocoon provides the summary of the table and ask the LLM to explain any expected duplication (\textit{ExpDuplicate}) beyond data errors, accompanied by a NL \textit{Thought}. (3) For the Semantic Review, if \textit{HasDuplicate} is true, Cocoon asks the LLM to evaluate whether the \textit{SampleDuplicate} is expected, based on \textit{ExpDuplicate} and \textit{Thought}.

\begin{example}
In a table for anonymous votes, with columns "vote\_date, candidate\_name," the Statistical Profile will alert \textit{HasDuplicate} with \textit{SampleDuplicate} (false positive). The Semantic Profile will consider the duplicates expected (\textit{ExpDuplicate}) as there are cases where multiple votes for the same person occur on the same day (\textit{Thought}). The Semantic Review will evaluate \textit{SampleDuplicate} and decide that this is expected, based on \textit{ExpDuplicate} and \textit{Thought}.
\end{example}

\subsubsection{Column Type}
There are two classes of types following Profiler~\cite{kandel2012profiler}: (1) primitive types for a single column, such as boolean, string, numeric (int, double), timestamp, date, and (2) higher-order types derived potentially from multiple columns, such geographic entities (currently including zip codes, FIPS codes, country names, US state names, and longitude and latitude coordinates), categories (as opposed to free text for string columns).

For primitive types, Cocoon processes each column as follows: (1) For the Statistical Profile, Cocoon uses the column type in DuckDB as its current type (\textit{CurrentType}) and takes a sample of 10 values (\textit{SampleValue}). (2) For the Semantic Profile, Cocoon provides a summary of the column and asks the LLM to identify any expected types (\textit{ExpType}) with a NL \textit{Thought}. (3) For the Semantic Review, it verifies the \textit{CurrentType} against the \textit{ExpType}. If they do not match, it examines the \textit{SampleValue} to evaluate if the \textit{CurrentType} is acceptable or if it indicates an error.

\begin{example}
For the Date column, values are represented in a non-standard format YYYYDDMM, like 20240412. The Statistical Profile specifies its \textit{CurrentType} as integer and detects no data quality errors (\textit{false negative}). The Semantic Profile, after being informed that the column's meaning is date, infers that the \textit{ExpType} should be date. The Semantic Review evaluates the discrepancy between \textit{CurrentType} and \textit{ExpType}. After inspecting \textit{SampleValue}, it concludes that it's a data error for not being recognized as date.
\end{example}

For higher-order types, Cocoon currently only classifies them; we leave error detection and the extension to other higher-order types in future work. Unlike Profiler, Cocoon allows for the derivation of higher-order types from multiple attributes. For example, longitude and latitude together denote geographic entities.
To classify columns, for each column group, Cocoon displays the first five rows from the projection of these columns, along with a summary for each column, and then asks the LLM to classify the higher-order types based on a list of columns.
After this classification, we treat all columns of a given higher-order type as a single column for subsequent error detection tasks.

\subsubsection{Unique key}
For each column, Cocoon operates as follows: (1) For the Statistical Profile, Cocoon issues an SQL query to compute the number of distinct values, divided by the total number of rows (\textit{UniqueRatio}), and takes a sample of at most 5 rows that share the same column value.
(2) For the Semantic Profile, Cocoon provides a summary of the table and column, and asks the LLM to explain whether this column is expected to be unique (\textit{ExpUnique}) with a NL \textit{Thought}. (3) For the Semantic Review, if \textit{ExpUnique} is true and \textit{UniqueRatio} $\neq1$, it evaluates \textit{SampleNonUnique} to decide whether these instances of non-uniqueness are acceptable.

\begin{example}
For a table with attributes "person\_id, address", the Statistical Profile shows that "person\_id" has a high \textit{UniqueRatio} of $95\%$, and highlights that the non-unique values are errors (\textit{false positive}). The Semantic Profile reviews the Semantic Context, noting that this table contains person residence addresses, and it is normal for some people to have multiple residences (\textit{ExpUnique} is false). The Semantic Review inspects the \textit{UniqueRatio}, \textit{SampleNonUnique}, and \textit{ExpUnique}, and concludes that the non-uniqueness is normal.
\end{example}

\subsubsection{Disguised Missing Values (DMV)}
For each column: (1) For the Statistical Profile, Cocoon applies FAHES~\cite{qahtan2018fahes} to detect candidate DMVs (\textit{CandidateDMV}). (2) For the Semantic Profile, given the Semantic Context of the column description, Cocoon asks the LLM to provide the expected DMVs for this column (\textit{PotentialDMV}) along with a NL \textit{Thought}. (3) For the Semantic Review, the LLM evaluates \textit{CandidateDMV}, \textit{PotentialDMV}, and \textit{Thought} to decide if \textit{CandidateDMV} are truly DMVs.

\begin{example}
For a table tracking user login times, the Statistical Profile identifies the date "January 1, 1970, at 00:00:00 UTC" as not \textit{CandidateDMV} due to its high frequency (\textit{false negative}). However, the Semantic Profile, understanding the context of Unix epoch time, flags this date as unrealistic for actual user logins, suggesting it is indeed a DMV (\textit{PotentialDMV}). The Semantic Review evaluates these findings, concluding that the date is a default placeholder, and therefore a DMV.
\end{example}

\subsubsection{Missing Values}
For each column, the process is as follows: (1) For the Statistical Profile, Cocoon computes the percentage of missing values (\textit{MissingPercentage}) and a sample of up to 5 rows with missing values (\textit{SampleMissing}). (2) For the Semantic Profile, based on the Semantic Context of table and column summary, Cocoon asks LLMs to determine whether missing values are allowable (\textit{AllowMissing}), and generates a rationale (\textit{Thought}). (3) For the Semantic Review, if \textit{MissingPercentage} $>0$, the acceptability of the \textit{SampleMissing} is assessed.

\subsubsection{Numeric Outliers}
For each numeric column: (1) In the Statistical Profile, Cocoon calculates the $0^{th}$ to $4^{th}$ quantiles (\textit{Quantile}) using DuckDB. (2) In the Semantic Profile, based on the semantic understanding of the table and column, Cocoon employs LLMs to identify the expected range of quantiles (\textit{ExpQuantile}) and to formulate a supporting explanation (\textit{Thought}). (3) During the Semantic Review, the alignment between the calculated quantiles (\textit{Quantile}) and the expected quantiles (\textit{ExpQuantile}) is examined by LLMs, who then evaluate the appropriateness of these quantiles.

\begin{example}
In the case of the Age column, the Statistical Profile identifies the $0^{th}$ to $4^{th}$ quantiles as $-2, -1, 10, 20, 90$ (\textit{Quantile}), marking $90$ as a potentially erroneous outlier. The Semantic Profile, recognizing the conceptual framework of age, anticipates the quantiles to be $0, 20, 40, 60, 130$ (\textit{ExpQuantile}). The Semantic Review then contrasts \textit{Quantile} with \textit{ExpQuantile}, acknowledging that while $90$ is an uncommon age, it is normal (\textit{false positive}), whereas ages $<0$ are considered anomalies since age cannot be negative (\textit{false negative}).
\end{example}

\subsubsection{String Outlier Analysis}
For each string column, the approach is as follows: (1) In the Statistical Profile, Cocoon leverages RAHA~\cite{mahdavi2019raha} to identify the prevalent regular expression pattern (\textit{RegexPattern}) and to select samples of 5 inliers and outliers for examination. (2) Within the Semantic Profile, leveraging the Semantic Context of the table and its columns, Cocoon uses LLMs to anticipate the values for the column (\textit{ExpStr}) and to generate an explanation (\textit{Thought}). (3) During the Semantic Review, the LLM evaluates the consistency between \textit{RegexPattern} and \textit{ExpStr}, examines \textit{SampleOutlier} to determine its acceptability, and \textit{SampleInlier} for possible errors.

\begin{example}
For SSN column, where all values begin with 999, the Statistical Profile identifies the RegexPattern of $999-\backslash d{2}-\backslash d{4}$ and observes no errors (\textit{false negative}). The Semantic Profile, understanding the meaning of SSN, offers legitimate examples such as $["210-58-9374", "539-24-5861", ...]$ (\textit{ExpStr}). The Semantic Review contrasts \textit{RegexPattern} with \textit{ExpStr} and notes that $999$ is an invalid area code, alerting that the entire column is erroneous.
\end{example}

\subsubsection{Missing Record}
For each categorical column: (1) In the Statistical Profile, Cocoon performs a count grouped by query for \textit{ValueFreq}. (2) For the Semantic Profile, considering the Semantic Context of the table and its columns, Cocoon employs LLMs to predict the value frequency (\textit{ValueFreq}) and generate a rationale (\textit{Thought}). (3) For the Semantic Review, the LLM assesses the consistency between \textit{ValueFreq} to identify potential missing records.

\begin{example}
For a sales table for Halloween costumes, the Statistical Profile calculates the count grouped by Month (\textit{ValueFreq}), and detects a pronounced bias towards October as missing records for the rest of the months (\textit{false positive}). The Semantic Profile, understanding the context of Halloween costumes, expects a high frequency during October (\textit{ExpFreq}). The Semantic Review contrasts \textit{ValueFreq} with \textit{ExpFreq} and concludes that the bias is normal.
\end{example}

\subsubsection{Human Feedback}
For all error types, Cocoon requests feedback from users by alerting them to the detected errors that need attention. The reasoning is also provided to help the user understand the nature of the error. Additionally, for Missing Records, we create visualizations to help users understand the distribution. We profile the column distribution using the following basic visualizations, prioritizing high-order over primitive data types:

\begin{itemize}
\item {\bf Histogram}: For numeric, timestamp, and date types.
\item {\bf Bar chart}: For categories and boolean types.
\item {\bf Map}: For geographic entities.
\end{itemize}

\subsubsection{Extension for Other Errors} Cocoon provides a common set of error types listed in \Cref{table:error_types} as a starting point. In the future, we aim to extend Cocoon to cover a more comprehensive set of data errors, such as those based on constraints~\cite{chu2013discovering}. The extension to other errors follows a similar pattern, where the appropriate statistical and semantic profiles must first be designed. Then, prompts need to be created to clearly articulate the semantic context, semantic profile, and semantic review tasks for LLMs.

\begin{figure}[h]
\centering
\begin{lstlisting}
{
  "Semantic Context": {
    "Table Summary": "The table contains medical and demographic records for  patients ...",
    "Attribute Hierarchy": {
      "Patient": {"Identification": [Id, SSN, DRIVERS...]},  
        ...},
    "Column Summary": {
      "id": "UUID (Universally Unique Identifier).", 
      "SSN": ...}
    },
  "Statistical Profile": {...},
  ...
}
\end{lstlisting}
\caption{Example JSON profile generated by Cocoon, used by downstream LLM applications to build prompts.}
\label{json}
\end{figure}

\subsection{Final Output}

Cocoon yields two types of outputs: (1) First, it provides analysts with a profile page that displays the final errors, as illustrated in \Cref{fig:teaser}, to assist in human understanding and facilitate future data cleaning tasks. It marks errors that are semantically erroneous with a red exclamation mark and explanation. It also displays errors that are only statistically, but not semantically, erroneous with a green tick and explanation. For each column group, it indicates how many alerts there are to help users navigate.
(2) Second, for future LLM applications in data tasks like data cleaning and Text-to-SQL, comprehending the significance of tables, columns, and potential data quality issues is a crucial initial step~\cite{huang2023data}. To support this, we supply JSON files that compile all the Semantic Context, Statistical Profile, Semantic Profile, and Semantic Review. An example of such a file is shown in \Cref{json}. Future LLM applications can utilize this file to select the relevant sections for constructing the prompts.

\section{User Study}

To evaluate whether Cocoon can accurately identify semantically meaningful data errors in real-world datasets for data analysts, we carried out a pilot study with two graduate students. One is from Columbia University's Climate School, and the other is from the Columbia University Irving Medical Center. Both participants extensively use data analytics daily. They provided datasets that are representative of their typical work. Due to the response times of LLMs ($\sim20$ minutes per dataset) and the limited availability of our participants, we generated the Cocoon profiles from the provided datasets offline. Afterwards, we interviewed the participants following a think-aloud protocol. They discussed (1) their views on the accuracy of Cocoon's profiling and (2) how Cocoon could be useful in their future data analysis tasks.

\subsection{Results}

\subsubsection{Climate} The first dataset collection includes crop production data from various locations throughout the year, provided by Participant 1 (P1). Cocoon flags errors, including duplicates, string outliers (such as typos and inconsistent representations), numerical outliers, missing values, and disguised missing values (for example, "-", "\#Value!"). Participant 1 notes, "Most of the errors identified by Cocoon are accurate. Its mistakes, however, are interesting". Upon requesting clarification from P1 about Cocoon's inaccuracies, it becomes evident that some errors have nuanced causes, rooted in domain-specific background knowledge. For instance, data drift issues arise when different notations are used across years (e.g., kilograms to pounds), leading to inconsistencies. Although Cocoon identifies these as errors, they are shifts in conventional practices; whether these are errors is ambiguous. Another situation where inconsistent representations are inevitable is due to uncertainty, such as a numerical field with a range $[min,max]$ instead of a single value, making the field non-numerical. Similarly, for a categorical field, while scales typically denote "small" or "medium", there is an occasional "small or medium". These issues underscore the necessity of human expertise in explaining the errors.

In interviews about how Cocoon could assist with future data analytics, participants express that Cocoon would be highly beneficial in identifying meaningful errors. "For context, last year, the cleaning was performed by an undergraduate who manually read through Excel sheets to find errors very similar to those flagged by Cocoon. But that was a slow process. We hope to use historical data, which is fraught with human errors. Sifting through 50 years of data for debugging is generally feasible (P1)". To fully leverage Cocoon, additional features are anticipated for comprehensive end-to-end cleaning in future work. "There will still be considerable work post-profiling. It would be helpful if the language model could assist in directly correcting typos (P1)".

\subsubsection{Medical} The second dataset encompasses OCT scan data for patients from Participant 2 (P2). Cocoon assists in identifying errors, including a significant number of duplicates, numerous numerical outliers in measurements, string outliers (inconsistent patterns), missing values, and disguised missing values (0). Participant 2 expresses surprise at the volume of data errors flagged by Cocoon. While most of the errors are acknowledged as intriguing, P2 finds it challenging to verify whether they are genuine errors without consulting the data provider. A specific type of error highlighted involves column misalignment, where one column's shift affects all subsequent columns. Although Cocoon reports these as individual errors for each column, identifying it as a single issue could enhance error reporting. This improvement could be considered in future developments by expanding the error categorization.

During discussions on how Cocoon could aid future data analytics, P2 shares, "We receive data from hospitals during one-hour meetings. During this brief period, we must review the data and inquire about any discrepancies, such as incorrect numbers or mismatched values. If we fail to address these issues immediately, we must wait until the next monthly meeting for clarification. Manually exploring the data thoroughly is impractical within such a limited timeframe. Cocoon will enable us to efficiently scan the data." This feedback underscores Cocoon's potential to streamline the data review process, allowing for more effective data quality control in time-constrained environments.

\section{Limitations and Future Works}

This work presents Cocoon, our approach to utilizing LLMs to provide semantically meaningful profiling. However, there are limitations that we plan to address in future work:

For the error type, we begin with the most common ones as listed in \Cref{table:error_types}, but the current number of types is still limited. Exploring how to semi-automatically crawl and build a repository that catalogs data errors~\cite{gudivada2017data}, potentially domain-specific, for challenges people face in the real world would be intriguing. Extending this to enhance the generality for real-world datasets is a goal.

For the profiling interface, our aim is to better design the interface to help users understand errors. We are currently developing, for each error type, a corresponding SQL query to help users trace back to understand the errors. This could be further facilitated by visualizations, as used by previous works~\cite{kandel2012profiler}. Furthermore, users also want to clean data within the same system. We aim to provide an interface to address error correction, combined with semantics.

Regarding downstream LLM applications, profiling and cleaning the data could greatly benefit them. We plan to experiment and demonstrate how the Cocoon profile could be used to improve data tasks such as Text-to-SQL~\cite{huang2023data} and transformation~\cite{huang2024relation}.

\pagebreak

\bibliographystyle{ACM-Reference-Format}
\bibliography{main}

\clearpage

\end{document}